# A Comparative Study of Retrieval Methods in Azure AI Search

Qiang Mao
Legal Technology & Data Analytics
Ankura Consulting Group, LLC
Washington, D.C. USA
qiang.mao@ankura.com

Han Qin
Legal Technology & Data Analytics
Ankura Consulting Group, LLC
Washington, D.C. USA
han.qin@ankura.com

Robert Neary
Legal Technology & Data Analytics
Ankura Consulting Group, LLC
Washington, D.C. USA
robert.neary@ankura.com

Charles Wang
Legal Technology & Data Analytics
Ankura Consulting Group, LLC
Washington, D.C. USA
charles.wang@ankura.com

Fusheng Wei
Legal Technology & Data Analytics
Ankura Consulting Group, LLC
Washington, D.C. USA
fusheng.wei@ankura.com

Jianping Zhang
Legal Technology & Data Analytics
Ankura Consulting Group, LLC
Washington, D.C. USA
jianping.zhang@ankura.com

Nathaniel Huber-Fliflet
Legal Technology & Data Analytics
Ankura Consulting Group, LLC
London, UK
nathaniel.huber-fliflet@ankura.com

***Abstract* – Increasingly, attorneys are interested in moving beyond keyword and semantic search to improve the efficiency of how they find key information during a document review task. Large language models (LLMs) are now seen as tools that attorneys can use to ask natural language questions of their data during document review to receive accurate and concise answers. This study evaluates retrieval strategies within Microsoft Azure's Retrieval-Augmented Generation (RAG) framework to identify effective approaches for Early Case Assessment (ECA) in eDiscovery. During ECA, legal teams analyze data at the outset of a matter to gain a general understanding of the data and attempt to determine key facts and risks before beginning full-scale review. In this paper, we compare the performance of Azure AI Search's keyword, semantic, vector, hybrid, and hybrid-semantic retrieval methods. We then present the accuracy, relevance, and consistency of each method's AI-generated responses. Legal practitioners can use the results of this study to enhance how they select RAG configurations in the future.**

## I. Introduction

Early Case Assessment (ECA) is the first phase of the eDiscovery process and focuses on rapidly reviewing key electronic data to assess the facts and risks of a legal matter. Effective ECA allows legal teams to gain an understanding of the data and make informed strategic decisions such as whether to settle, litigate, or attempt to narrow the issues. ECA involves identifying, collecting, and analyzing electronically stored information (ESI) to surface key facts, scope the issues, and defensibly reduce non-relevant data before full review. This process helps legal teams estimate potential costs, narrow downstream workflows, and assess likely case outcomes.

Recently, legal teams have started using Generative AI (GenAI) to analyze large volumes of data more efficiently when compared to traditional keyword or semantic search methods. Law firms increasingly expect tools that can support GenAI enabled ECA workflows such as case analysis, issue identification, document classification, and document summarization through natural-language interaction with their data.

Retrieval-Augmented Generation (RAG) is a prominent approach that enhances the capabilities of large language models (LLMs) by incorporating external knowledge retrieved from indexed data sources and grounding the model's responses in that information. RAG combines two key components: a retriever that fetches relevant documents and a generator that synthesizes responses based on both the prompt and retrieved content.

However, the effectiveness of these systems depends heavily on the retrieval component that selects which documents are provided to the model. Unlike keyword search, where the behavior is well understood, the accuracy of vector, semantic, hybrid, and reranked retrieval methods in real eDiscovery-like conditions remains unclear. Their performance can vary widely depending on the structure of the query, the vocabulary used, and the distribution of evidence across a dataset.

This uncertainty creates real risk for legal workflows. If a retrieval method fails to surface the right documents, a GenAI system may return incomplete or incorrect answers, undermining defensibility in areas such as privilege identification, issue analysis, and early case assessment. Legal teams need empirical evidence showing how these retrieval methods behave so they can choose configurations that maximize recall, reduce error rates, and support accurate, defensible analysis.

This paper evaluates RAG implementations on Microsoft Azure, using Azure AI Search as the retriever and Azure OpenAI as the generator. We evaluate how different retrieval methods affect the accuracy and relevance of generated outputs.

## II. Background and Azure RAG Architecture

Retrieval-Augmented Generation (RAG) is a framework that enhances LLMs by incorporating retrieved documents into the generation process. Instead of relying solely on the model's internal knowledge, RAG retrieves relevant chunks of external data and feeds them into the LLM alongside the user prompt. This approach can improve factual accuracy, allow domain-specific customization, and reduce hallucinations by grounding responses in retrieved content.

The Azure RAG pipeline includes data ingestion, preprocessing, indexing, retrieval and ranking, generation, and response parsing. Each step plays a critical role in ensuring the relevance and accuracy of the final output.

Azure AI Search supports several retrieval strategies: Keyword, Keyword-Semantic, Vector, Hybrid, and Hybrid-Semantic [1].

- *Keyword search* uses a traditional lexical approach to match query terms against indexed text and rank documents based on term frequency and inverse document frequency.
- *Keyword-Semantic search* begins with keyword retrieval and then applies a semantic scoring model to improve the ordering of those keyword results.
- *Vector search* embeds both the query and the indexed text into a numerical vector space and retrieves documents based on cosine similarity, enabling conceptual matching even when exact terms do not appear.
- *Hybrid search* combines keyword and vector retrieval by merging the top results from both methods to capture exact matches and semantic matches simultaneously.
- *Hybrid-Semantic search* extends hybrid retrieval by applying a semantic reranker that reorders the merged results based on deeper contextual similarity to the query.

Each method offers a trade-off between lexical matching, semantic understanding, and computational cost. Semantic reranking reorders retrieved documents based on similarity to the query and typically improves relevance but at additional computational cost.

*Table 1 Azure AI Search Retrieval Methods*

| **Retrieval Method** | **Lexical Match** | **Semantic Understanding** | **Vector Similarity** | **Reranking** | **Cost Impact** |
|---|---|---|---|---|---|
| Keyword | ✓ | ✗ | ✗ | ✗ | Low |
| Keyword-Semantic | ✓ | ✓ | ✗ | ✓ | Medium |
| Vector | ✗ | ✗ | ✓ | ✗ | Low |
| Hybrid | ✓ | ✗ | ✓ | ✗ | Low |
| Hybrid-Semantic | ✓ | ✓ | ✓ | ✓ | High |

Table 1 summarizes the key characteristics of each retrieval strategy supported by Azure AI Search. It compares their reliance on lexical matching, semantic interpretation, vector similarity, and reranking, as well as the relative cost impact associated with each method.

## III. Experiments

We used the Jeb Bush email dataset comprised of over 290,000 text documents. It's publicly available and considered comparable to the data present in most modern eDiscovery projects. After chunking each document into 2,000-token segments with a 500-token overlap, the dataset yielded 491,482 chunks. We embedded each chuck using the 'text-embedding-ada-002' model with a vector dimension of 1536. We used 'gpt-4.1-mini' as our LLM. All LLM generation was performed with the temperature set to zero to provide deterministic outputs and isolate the impact of retrieval differences. For each query, the top 50 chunks were retrieved, and the top 5 were sent to the LLM.

We prepared the following prompts to demonstrate the effects of different retrieval methods:

*Table 2 The Prompts*

| | |
|---|---|
| P1 | *What is case no. TO 98-103033, 32?* |
| P2 | *What were the projected volumes of orange production in Florida, and how would they affect Florida's economy?* |
| P3 | *Was D&B Television doing business in West Palm Beach?* |
| P4 | *Who is T-Squred?* |
| P5 | *Who is 'T-Squred'?* |
| P6 | *Find emails discussing differences between Jeb Bush and George Bush's position on climate change and environmental policy using natural language concepts, even if those exact terms aren't used.* |
| P7 | *What were Jeb Bush's views on the H-1B visa program during his tenure as governor of Florida?* |
| P8 | *What were Jeb Bush's views on the H-1B visa program during his tenure as governor of Florida? Please ensure that all responses reflect Jeb Bush's own views specifically, not those of other individuals or political figures.* |
| P9 | *What were Jeb Bush's views on the H1B visa program during his tenure as governor of Florida? Please ensure that all responses reflect Jeb Bush's own views specifically, not those of other individuals or political figures, including President Bush.* |

We evaluated each retrieval method on relevance, factual accuracy, and diversity of generated responses. These assessments were qualitative and based on manual review of retrieved chunks and model responses.

## IV. Results and Discussion

In the following, we discuss the RAG responses to prompts using different retrieval methods.

**Prompt P1**: The case number is a very specific alphanumeric string that does not contain ordinary words. Since the case number is mentioned in the emails, it becomes a keyword indexed by AI search. Unsurprisingly, methods utilizing keyword search enable RAG to respond with correct references to the emails. Conversely, the vector embedding of the prompt does not embed the alphanumeric case number as a meaningful semantic feature, causing the vector retrieval method to fail to locate the emails.

**Prompt P2**: This scenario is somewhat the opposite of P1. The vector retrieval method successfully identified emails containing specific orange product forecast numbers, whereas the keyword method failed to locate such forecasts. Additionally, the keyword semantic method did not identify them either, even though both methods partially addressed the economic impact portion of the query. The failure of the semantic method suggests that the keyword search did not return relevant information within its 50 results; because semantic reranking only reorders keyword search results, it cannot recover documents that keyword retrieval does not surface.

**Prompt P3**: An email message indicates that the company D&B is conducting business in South Florida. This information was overlooked by both the keyword and vector methods, possibly due to the lack of semantic association between "West Palm Beach" and "South Florida" in the embedding space and the original ranking of the keyword search. After semantic reranking, this email was moved to the top and subsequently picked up by the LLM step.

**Prompt P4**: T-Squired is the nickname of an individual referenced in the dataset. Due to its unique meaning, similar to the outcome in P3, both simple and vector retrieval failed to extract the correct information. However, the semantic method was successful.

**Prompt P5**: By enclosing T-Squared in quotes, the keyword method now locates the correct information, while the vector method continues to fail.

**Prompt P6**: This example shows that while keyword, vector, and keyword semantic methods all stated that the requested information is unavailable, the hybrid and hybrid-semantic methods cited emails from others revealing a "nuanced difference." Upon reviewing the cited emails, the responses were partially relevant but did not directly answer the question.

**Prompt P7**: While keyword, vector, and hybrid methods all indicated they could not find information about the governor's stance on H1-B, the keyword semantic and hybrid semantic methods erroneously attributed others' positions to Jeb Bush, as shown in the citations. Upon examining the citations in the LLM response, it appears that re-ranking elevated a text chunk to the top retrieval results, which was part of speech given by President George Bush. Because the LLM lacks source awareness, it inferred that the retrieved text represented Jeb Bush's views. This illustrates that while semantic re-ranking can aid in selecting top documents for the LLM, it can also lead to unintended consequences.

**Prompt P8**: This prompt repeats the question from P7, requesting the LLM to verify the answer. The same text chunk was selected, and because the same chunk was retrieved, the LLM reproduced the same attribution error.

**Prompt P9**: Finally, noting that the citations from P7 and P8 mistakenly attributed George Bush's views on H1-B to Jeb Bush, the prompt was further refined. This time, the erroneous chunk no longer appeared in the top retrieval results, and no views were attributed to Jeb Bush.

We observed that Keyword retrieval delivered faster results faster than other methods but missed semantically relevant material. Vector retrieval captured semantic similarity but occasionally returned off-topic chunks for infrequent keywords. Hybrid Semantic produced the most consistently relevant results but at a higher computational cost.

Appendix A provides the full prompt set and the generated outputs for each retrieval method. These example results illustrate the retrieval behavior directly, provide visibility to the observations reported in this work, and illustrate how specific retrieval methods influenced the LLM's grounded responses.

These observations are a result of our specific experimental configuration choices. Adjusting retrieval settings, chunking parameters, or model versions may yield different outcomes. Nonetheless, these cases demonstrate core retrieval behaviors that practitioners should expect to see when using certain RAG implementations.

## V. Conclusions

This study demonstrates that the choice of retrieval method in Azure RAG has a direct impact on the accuracy, relevance, and defensibility of AI-generated responses. Our prompt-level analysis shows that no single method performs consistently across all scenarios; instead, each method succeeds or fails based on the nature of the query and the distribution of information in the dataset. Many errors attributed to "LLM hallucination" were rooted in retrieval method design deficiencies, underscoring that retrieval quality is the primary determinant of downstream model performance.

Several patterns emerged across the experiments. Keyword-based retrieval remains essential for structured, non-semantic data, such as case numbers, IDs, and nicknames, while vector retrieval excels at locating conceptually related content but struggles with domain-specific or infrequent terms. Semantic reranking improved recall in many cases but also elevated misleading or tangential text when the initial results lacked relevant material. These have clear implications for legal workflows, particularly in high-stakes contexts such as privilege identification, issue analysis, and ECA. Retrieval errors, such as elevating misattributed information or tangential text, can lead the LLM to generate confident but incorrect answers. Without transparent citations and auditable retrieval paths, these failures undermine defensibility and increase the risk of relying on incorrect information in litigation or regulatory matters. The variability across retrieval strategies reinforces the need for multi-method retrieval options, calibrated reranking, and clear visibility into what content the system uses to support its answers.

For eDiscovery workflows, these findings reinforce the importance of giving lawyers a clear understanding of the implications of RAG method selection, surfacing retrieval scores, and providing visibility into the source of the retrieved content. They also show that prompt phrasing has a strong impact on retrieval behavior, meaning that user input, not only system configuration, affects output performance.

## VI. Future Works

Future work will include visualizing retrieval and reranking behavior to help users understand why specific documents surfaced, incorporating and testing feedback loops to inform a system to adaptively select more efficient retrieval methods over time, and benchmarking retrieval strategies across larger and more diverse datasets to validate performance in real-world legal scenarios. Another area of exploration will use an agentic system to evaluate each query and select the retrieval strategy (e.g., keyword, vector, hybrid, semantic) that would likely produce the most accurate results. Such systems could also review and verify the LLM-generated result. These advancements could help practitioners deploy RAG systems that are more accurate, predictable, and defensible in high-stakes environments.

## Appendix A. RAG Experiment Output

| Prompt | Keyword | Vector | Keyword Semantic | Hybrid | Hybrid Semantic |
|---|---|---|---|---|---|
| what is case no. TO 98-103033, 32 | Case no. TO 98-103033, 32 refers to a criminal court case in Orange County involving Angelo Albenzi. The case was scheduled for a pretrial conference (PTC) on August 3, with the trial set to begin during the court period starting August 9. Angelo Albenzi mentioned that during the trial phase, he would demand a full transcript to be supplied to him for the Florida Department of Law Enforcement (FDLE) [doc1]. Additionally, it is noted that this case never went to trial, and the defendant did not deny driving the vehicle involved [doc2].<br><br>Citations:<br>[doc1] : JEB0020-027057<br>[doc2] : JEB0045-032981<br><br>Top chunks sent to LLM:<br>JEB0020-027057<br>JEB0045-032981<br>JEB1086-036557<br>JEB0895-180574<br>JEB1320-206709 | The retrieved documents do not contain any information about case number TO 98-103033, 32.<br><br>The requested information is not available in the retrieved data. Please try another query or topic.<br><br>Citations:<br><br>Top chunks sent to LLM:<br>JEB2690-155704<br>JEB0045-032981 | Case no. TO 98-103033, 32 refers to a criminal court case in Orange County involving Angelo Albenzi. He was due in court on August 3 for a pretrial conference (PTC) to set the trial beginning in the court period of August 9. The case involves issues related to a traffic stop and matters concerning a suspended driver's license and unpaid citations. Angelo Albenzi mentioned that the defendant in this case never denied driving the vehicle, and he planned to demand a full transcript for the FDLE (Florida Department of Law Enforcement) [doc1][doc2].<br><br>Citations:<br>[doc1] : JEB0045-032981<br>[doc2] : JEB0020-027057<br><br>Top chunks sent to LLM:<br>JEB0045-032981<br>JEB0020-027057<br>JEB9226-218265<br>JEB10179-203386<br>JEB10179-203386 | Case no. TO 98-103033, 32 refers to a criminal court case in Orange County involving Angelo Albenzi. The case was scheduled for a pretrial conference (PTC) on August 3, with the trial set to begin during the court period starting August 9. Angelo Albenzi mentioned that during the trial phase, he would demand a full transcript to be supplied to the Florida Department of Law Enforcement (FDLE) in Orlando. The case appears to involve issues related to law enforcement and legal proceedings, with Albenzi expressing concerns about enforcement of Florida laws and indicating potential federal litigation threats [doc1][doc2].<br><br>Citations:<br>[doc1] : JEB0020-027057<br>[doc2] : JEB0045-032981<br><br>Top chunks sent to LLM:<br>JEB0020-027057<br>JEB0045-032981<br>JEB0895-180574<br>JEB2690-155704<br>JEB0045-032981 | Case no. TO 98-103033, 32 refers to a criminal court case in Orange County involving Angelo Albenzi. The case was scheduled for a pretrial conference (PTC) on August 3, with the trial set to begin during the court period of August 9. Angelo Albenzi mentioned that during the trial phase, he would demand a full transcript to be supplied to him for the Florida Department of Law Enforcement (FDLE). He also expressed a desire for an FDLE employee to be present during the proceedings due to concerns about law enforcement's ability to enforce Florida laws [doc2].<br><br>Citations:<br>[doc2] : JEB0020-027057<br><br>Top chunks sent to LLM:<br>JEB0045-032981<br>JEB0020-027057<br>JEB0045-032981<br>JEB0045-032981<br>JEB0797-103033 |
| What were the projected volumes of orange production in Florida, and how would they affect Florida's economy? | The retrieved documents do not provide specific projected volumes of orange production in Florida. However, they do highlight the significant economic impact of Florida's citrus industry. The unsubsidized Florida citrus industry contributes $9.1 billion to the state's economy and employs over 90,000 people, making it typically the state's second-largest industry after tourism. The industry also supports many related businesses such as equipment, chemical, fertilizer suppliers, and manufacturers. Additionally, the current citrus tariff helps maintain a level playing field between Florida and Brazil in the orange juice market, protecting Florida's industry from being severely crippled by competition. A reduction or elimination of this tariff could lead to a substantial decline in grower revenues and income, estimated at about $500 per acre annually, which would represent approximately a 90 percent reduction in average per acre income for 1999-2000. This would negatively affect Florida's economy by weakening its second-largest industry and increasing market control by Brazilian processors, potentially offering no savings to U.S. consumers [doc1].<br><br>Citations:<br>[doc1] : JEB0929-113131<br><br>Top chunks sent to LLM:<br>JEB0929-113131<br>JEB1706-254227<br>JEB1706-254227<br>JEB1708-254229<br>JEB1707-254228 | The projected volumes of orange production in Florida varied across different seasons and reports:<br><br>- For the 2006-07 season, the Florida Department of Citrus budget was based on a statewide orange production of 180 million boxes, which was an increase from the U.S. Department of Agriculture's projection of 153 million boxes for that season [doc1].<br>- In the 2006-07 season, the U.S. Department of Agriculture predicted 135 million boxes of oranges would be picked, down from 148 million boxes the previous year, marking the worst crop in more than 15 years [doc2].<br>- For an earlier season, the USDA estimated Florida growers would pick a record 252 million 90-pound boxes, a 24 percent increase from the 2002-03 season [doc3].<br><br>Regarding the economic impact:<br><br>- The Florida citrus industry has an economic impact of $9.1 billion to the state and employs over 90,000 people, making it typically the state's second-largest industry after tourism. This impact extends beyond growers to equipment, chemical, fertilizer suppliers, manufacturers, and the local economy [doc4].<br>- A decline in grower revenues and income due to reduced production or tariff changes could severely affect the industry, with potential income reductions of about $500 per acre, representing a significant loss [doc4].<br>- Oversupply and falling consumption have led Florida growers to seek government intervention, such as requesting the USDA to buy 90 million gallons of orange | The retrieved documents do not provide specific projected volumes of orange production in Florida. However, they highlight the significant economic impact of Florida's citrus industry, stating that the unsubsidized Florida citrus industry has an economic impact of $9.1 billion to the state and employs over 90,000 people, making it typically the state's second-largest industry after tourism [doc1]. The documents also emphasize that Florida and Brazil together produce roughly 90 percent of the world's orange juice, with Florida consuming 90 percent of its production domestically and exporting 10 percent [doc1][doc2]. The tariff on orange juice imports helps protect Florida's citrus industry from Brazilian competition, and removing or reducing this tariff could severely harm Florida's citrus growers by reducing their revenues significantly [doc1][doc2].<br><br>In summary, while exact projected volumes are not mentioned, the Florida orange production is a major economic driver for the state, and changes in trade policies affecting production or export could have substantial economic consequences .<br><br>Citations:<br>[doc1] : JEB0929-113131<br>[doc2] : JEB0929-113131<br><br>Top chunks sent to LLM:<br>JEB0929-113131<br>JEB0929-113131<br>JEB0971-098751<br>JEB1030-098810<br>JEB3178-277403 | The projected volumes of orange production in Florida varied across different seasons. For the 2006-07 season, the budget was based on a statewide orange production of 180 million boxes, which was an increase from the U.S. Department of Agriculture's projection of 153 million boxes for that season [doc1]. In the 2003-04 season, Florida growers were projected to produce 246 million boxes of oranges [doc4]. Earlier, the USDA estimated a record 252 million 90-pound boxes for a season around 2002-03, which was a 24 percent increase from the previous season [doc5]. However, the 2006-07 season was also predicted to be the worst crop in more than 15 years with a forecast of 135 million boxes, down from 148 million boxes the previous year [doc3].<br><br>The economic impact of Florida's citrus industry is significant. It has an economic impact of $9.1 billion to the state and employs over 90,000 people, making it typically the state's second-largest industry after tourism [doc2]. The industry also supports many related businesses such as equipment, chemical, fertilizer suppliers, and manufacturers, contributing broadly to the local economy [doc2]. The tariff on orange juice imports from Brazil helps protect this industry by preventing cheaper imports from flooding the U.S. market, which could severely cripple Florida's citrus industry and reduce grower revenues substantially [doc2][doc4]. Additionally, the Florida Department of Citrus has sought federal support to purchase orange juice to help balance production and consumption, which would help stabilize prices and support growers [doc5].<br><br>In summary, Florida's orange production volumes have | Florida's orange production projections varied over different years. In one instance, the USDA and National Agricultural Statistics Board predicted Florida would produce 226 million boxes of oranges in a year, which would be the state's third largest orange crop on record [doc1]. Another forecast by Dr. Elizabeth Steger of Citrus Consulting International projected a record 247 million boxes of oranges, surpassing the previous record of 240 million boxes produced in 1997-98 [doc2]. For the 2003-04 season, Florida growers were projected to produce 246 million boxes of oranges [doc5]. However, in the 2006-07 season, the U.S. Department of Agriculture predicted a significant drop to 135 million boxes, the worst crop in more than 15 years, due to hurricanes, drought, and disease [doc3].<br><br>Economically, the Florida citrus industry is highly significant, with an economic impact of $9.1 billion to the state and employing over 90,000 people, making it typically the state's second largest industry after tourism [doc4]. The industry supports many related businesses, including equipment, chemical, fertilizer suppliers, and local economies [doc4]. The federal tariff on orange juice imports from Brazil helps protect Florida growers by preventing cheaper Brazilian juice from flooding the U.S. market, which would severely harm Florida's citrus industry and reduce grower revenues by about $500 per acre annually, representing a 90 percent reduction in average income per acre [doc4][doc5]. Thus, the volume of orange production directly influences Florida's economy through employment, industry revenue, and related sectors .<br><br>Citations: |
| Was D&B Television doing business in West Palm Beach? | The retrieved documents do not contain any information about D&B Television doing business in West Palm Beach. Therefore, the requested information is not available in the retrieved data. Please try another query or topic.<br><br>Citations:<br><br>Top chunks sent to LLM:<br>JEB0613-133465<br>JEB2581-173365<br>JEB2741-173525<br>JEB3045-173829<br>JEB0234-198609 | The retrieved documents do not contain any information about D&B Television doing business in West Palm Beach. Therefore, the requested information is not available in the retrieved data. Please try another query or topic.<br><br>Citations:<br><br>Top chunks sent to LLM:<br>JEB0686-231594<br>JEB1041-220080<br>JEB0702-239220<br>JEB0194-009161<br>JEB0173-121652 | Yes, D&B Television and Video Productions Inc. was doing business in West Palm Beach. They confirmed that they were operating normally in South Florida and were available for productions throughout the Eastern seaboard, including Miami, which is near West Palm Beach. They also mentioned that their crews could be available as local crews, which implies service availability in the region [doc1].<br><br>Citations:<br>[doc1] : JEB0031-000032<br><br>Top chunks sent to LLM:<br>JEB0031-000032<br>JEB0142-020613<br>JEB2314-197424<br>JEB1220-028609<br>JEB1217-028606 | The retrieved documents do not contain any information about D&B Television doing business in West Palm Beach. Therefore, the requested information is not available in the retrieved data. Please try another query or topic.<br><br>Citations:<br><br>Top chunks sent to LLM:<br>JEB0613-133465<br>JEB2581-173365<br>JEB1647-275872<br>JEB0702-239220<br>JEB1041-220080 | Yes, D&B Television and Video Productions Inc. was doing business in South Florida, which includes West Palm Beach. They reassured that they were operating as normal in South Florida and were available for productions throughout the Eastern seaboard, including Miami, which is near West Palm Beach [doc1].<br><br>Citations:<br>[doc1] : JEB0031-000032<br><br>Top chunks sent to LLM:<br>JEB0031-000032<br>JEB0142-020613<br>JEB3741-226037<br>JEB2314-197424<br>JEB0864-029695 |

| Prompt | Keyword | Vector | Keyword Semantic | Hybrid | Hybrid Semantic |
|---|---|---|---|---|---|
| Who is T-Squred? | The requested information is not found in the retrieved data. Please try another query or topic.<br><br>Citations:<br><br>Top chunks sent to LLM: | The retrieved documents do not contain any information about "T-Squred."<br><br>The requested information is not available in the retrieved data. Please try another query or topic.<br><br>Citations:<br><br>Top chunks sent to LLM:<br>JEB0951-253472<br>JEB0784-033720<br>JEB2826-138882<br>JEB0755-006592<br>JEB10410-203641 | Based on the retrieved documents, "T-Squared" appears to be a nickname used for Teresa Tinker. She is involved in coordinating and communicating on various projects, such as briefing officials, managing events, and working with consultants on initiatives like true cost accounting and pilot community selections [doc4][doc5]. The nickname "T-Squared the Magnificent" is used informally in communications, including by Jeb Bush [doc2][doc4].<br><br>Citations:<br>[doc2] : JEB2129-076629<br>[doc4] : JEB1355-008109<br>[doc5] : JEB1019-121477<br><br>Top chunks sent to LLM:<br>JEB2136-076636<br>JEB2129-076629<br>JEB4085-097682<br>JEB1355-008109<br>JEB1019-121477 | The retrieved documents do not contain any information about "T-Squred."<br><br>The requested information is not available in the retrieved data. Please try another query or topic.<br><br>Citations:<br><br>Top chunks sent to LLM:<br>JEB0951-253472<br>JEB0031-155872<br>JEB0394-190596<br>JEB0776-058639<br>JEB0784-033720 | Based on the retrieved documents, "T-Squared" appears to be a nickname used for Teresa Tinker. She is involved in coordinating and communicating on various projects, such as briefing officials, working with consultants, and organizing pilot communities for initiatives like true cost accounting and growth management [doc4][doc5]. The nickname "T-Squared the Magnificent" is used informally in the communications, including by Jeb Bush [doc2][doc4].<br><br>Citations:<br>[doc2] : JEB2129-076629<br>[doc4] : JEB1355-008109<br>[doc5] : JEB1019-121477<br><br>Top chunks sent to LLM:<br>JEB2136-076636<br>JEB2129-076629<br>JEB4085-097682<br>JEB1355-008109<br>JEB1019-121477 |
| Who is 'T-Squred'? | Based on the retrieved documents, "T-Squared" appears to be a nickname for Teresa Tinker. She is involved in various communications related to government and planning matters, such as the true cost accounting work group and growth management discussions [doc2][doc5]. The nickname "T-Squared the Magnificent" is mentioned as a very nice nickname that is likely to stick, indicating familiarity and a positive regard for Teresa Tinker [doc4].<br><br>Citations:<br>[doc2] : JEB1355-008109<br>[doc4] : JEB2129-076629<br>[doc5] : JEB1019-121477<br><br>Top chunks sent to LLM:<br>JEB2089-079295<br>JEB1355-008109<br>JEB3730-230074<br>JEB2129-076629<br>JEB1019-121477 | The retrieved documents do not contain any information about "T-Squred."<br><br>The requested information is not available in the retrieved data. Please try another query or topic.<br><br>Citations:<br><br>Top chunks sent to LLM:<br>JEB0003-182952<br>JEB0951-253472<br>JEB0394-190596<br>JEB0426-010815<br>JEB1235-244567 | "T-Squared" is a nickname for Teresa Tinker, who appears to be involved in various governmental or administrative activities, including coordinating work groups and consulting on projects such as true cost accounting and development issues [doc3][doc4][doc5]. The nickname is used informally in communications with Jeb Bush and others [doc1][doc2][doc3][doc4].<br><br>Citations:<br>[doc1] : JEB2136-076636<br>[doc2] : JEB2129-076629<br>[doc3] : JEB1355-008109<br>[doc4] : JEB1019-121477<br>[doc5] : JEB2755-229099<br><br>Top chunks sent to LLM:<br>JEB2136-076636<br>JEB2129-076629<br>JEB1355-008109<br>JEB1019-121477<br>JEB2755-229099 | Based on the retrieved documents, "T-Squared" appears to be a nickname or informal reference used for Teresa Tinker, as indicated in the email from Teresa Tinker where the message is addressed to Jeb Bush and others, and Jeb Bush responds with "Looks good to me t squared!" [doc5].<br><br>Citations:<br>[doc5] : JEB1355-008109<br><br>Top chunks sent to LLM:<br>JEB0394-190596<br>JEB2089-079295<br>JEB0003-182952<br>JEB0951-253472<br>JEB1355-008109 | "T-Squared" is a nickname for Teresa Tinker, who appears to be involved in various governmental or administrative activities, including coordinating work groups and consulting on projects such as true cost accounting and development issues [doc3][doc4][doc5]. The nickname is used informally in communications with Jeb Bush and others [doc1][doc2][doc3][doc4].<br><br>Citations:<br>[doc1] : JEB2136-076636<br>[doc2] : JEB2129-076629<br>[doc3] : JEB1355-008109<br>[doc4] : JEB1019-121477<br>[doc5] : JEB2755-229099<br><br>Top chunks sent to LLM:<br>JEB2136-076636<br>JEB2129-076629<br>JEB1355-008109<br>JEB1019-121477<br>JEB2755-229099 |

| Prompt | Keyword | Vector | Keyword Semantic | Hybrid | Hybrid Semantic |
|---|---|---|---|---|---|
| Find emails discussing differences between Jeb Bush and George Bush's position on climate change and environmental policy using natural language concepts, even if those exact terms aren't used | The retrieved documents do not contain any emails or content discussing differences between Jeb Bush and George Bush specifically on climate change or environmental policy using natural language concepts. The documents include emails about environmental concerns related to Florida Administrative Code changes and some commentary on Jeb Bush's political style and ideology compared to his brother George Bush, but none address their positions on climate change or environmental policy directly or comparatively.<br><br>Therefore, the requested information is not available in the retrieved data. Please try another query or topic.<br><br>Citations:<br><br>Top chunks sent to LLM:<br>JEB0178-170962<br>JEB0991-275216<br>JEB0840-199215<br>JEB0184-170968<br>JEB3667-277892 | The retrieved documents do not contain any emails explicitly discussing differences between Jeb Bush and George Bush's positions on climate change or environmental policy. The emails mention Jeb Bush's involvement in environmental issues and concerns about fossil fuels and climate change, but there is no direct comparison or discussion of differences between Jeb Bush and George Bush on these topics in the provided texts.<br><br>Therefore, the requested information is not available in the retrieved data. Please try another query or topic.<br><br>Citations:<br><br>Top chunks sent to LLM:<br>JEB0333-098113<br>JEB0156-058019<br>JEB1503-086372<br>JEB0313-066812<br>JEB0520-006357 | The retrieved documents do not contain any emails or content discussing differences between Jeb Bush and George Bush specifically on climate change or environmental policy. The only relevant content about Jeb Bush and George Bush compares their general political styles and ideologies but does not mention climate change or environmental policy explicitly [doc3]. Other documents focus on environmental issues in Florida and related political actions but do not compare the two Bushes on these topics [doc1, doc2, doc4, doc5].<br><br>Therefore, no emails discussing differences between Jeb Bush and George Bush on climate change or environmental policy are found in the retrieved data.<br><br>Citations:<br>[doc3] : JEB0991-275216<br><br>Top chunks sent to LLM:<br>JEB0232-171016<br>JEB1013-119230<br>JEB0991-275216<br>JEB0796-004895<br>JEB0090-170874 | The following emails discuss differences between Jeb Bush and George Bush's positions on climate change and environmental policy using natural language concepts:<br><br>1. In an email from Lanny E Smith to Jeb Bush, Lanny expresses concern about President George Bush's policies, particularly on drilling off the coast of Florida and the administration's "oil is our savior" thinking. Lanny urges Jeb Bush to take a stand against his brother's policy and to educate himself on the ramifications of climate change, emphasizing the need for leadership to develop clean energy sources and protect the environment [doc2].<br><br>2. Another email from William Eagan to Jeb Bush highlights anger over President George Bush's environmental policies, noting that even some Republicans are alarmed by the loss of wilderness and species extinctions due to insufficient protective measures. The email contrasts George Bush's "unconservation actions" with traditional conservative conservation ethics and expresses hope in Jeb Bush and other GOP officeholders who recognize the intrinsic value of nature [doc3][doc5].<br><br>3. An email from Susan (last name not provided) discusses the political and public perception challenges related to George Bush's environmental stance, particularly on drilling and arsenic regulations. It suggests that George Bush's approach to environmental issues may hurt him politically, especially in Florida, and implies that Jeb Bush may face pressure to oppose his brother's policies to protect the state [doc4]. | Two emails discuss differences between Jeb Bush and George Bush regarding climate change and environmental policy using natural language concepts:<br><br>1. In an email from Lanny E Smith to Jeb Bush dated January 29, 2001, Lanny urges Jeb to take a stand against his brother President George Bush's policies, especially on drilling off the coast of Florida. Lanny expresses concern about global warming caused by fossil fuels and calls for leadership to develop clean energy sources. This email highlights a divergence where Jeb is encouraged to oppose George's administration's oil-focused policies for environmental reasons [doc2].<br><br>2. Another email exchange from May 22, 2004, between William Eagan and Jeb Bush references a publication from Republicans for Environmental Protection. William expresses frustration with President George Bush's environmental policies, which some conservatives find unconservative and harmful to wild lands. Jeb responds somewhat defensively, listing Florida's environmental achievements under his governance but acknowledges the criticism of George Bush's policies. This exchange reflects ideological and policy differences on environmental conservation between the two brothers [doc3][doc4].<br><br>These emails illustrate that while Jeb Bush has taken some conservation actions and is seen as more environmentally conscious, there is tension and disagreement with George Bush's administration's approach to climate and environmental issues. The language used frames the differences in terms of leadership, policy priorities, and conservation ethics |
| What were Jeb Bush's views on the H1B visa program during his tenure as governor of Florida? | The retrieved documents do not contain any information about Jeb Bush's views on the H1B visa program during his tenure as governor of Florida. They include an email exchange regarding a specific H1B visa case but do not provide Jeb Bush's general views or policies on the H1B visa program.<br><br>The requested information is not available in the retrieved data. Please try another query or topic.<br><br>Citations:<br><br>Top chunks sent to LLM:<br>JEB1749-287501<br>JEB1750-287502<br>JEB1311-157152<br>JEB3009-263178<br>JEB3052-263221 | The requested information is not available in the retrieved data. Please try another query or topic.<br><br>Citations:<br><br>Top chunks sent to LLM:<br>JEB1161-191363<br>JEB0757-193353<br>JEB0333-098113<br>JEB0156-058019<br>JEB1503-086372 | Jeb Bush expressed support for increasing the cap on H1B visas during his tenure. He believed that allowing more skilled foreign workers to come to the U.S. under the H1B visa program was important for filling jobs that were difficult to fill domestically, helping companies like 3M remain competitive and maintain a strong job base. Bush emphasized the need for America to be flexible and innovative by welcoming talented individuals with the necessary skill sets, and he called on Congress to be realistic and reasonable in raising the H1B visa cap to support the country's leadership in innovation and economic strength [doc1].<br><br>Citations:<br>[doc1] : JEB0267-260436<br><br>Top chunks sent to LLM:<br>JEB0267-260436<br>JEB0221-146491<br>JEB1125-280030<br>JEB3344-225640<br>JEB1870-131535 | The retrieved documents do not provide specific information about Jeb Bush's views on the H1B visa program during his tenure as governor of Florida. There is a mention of a case involving an individual on an H1B visa in correspondence with Jeb Bush, but no explicit statement of his views on the H1B visa program itself is included in the documents [doc3].<br><br>Therefore, the requested information is not available in the retrieved data. Please try another query or topic.<br><br>Citations:<br>[doc3] : JEB1750-287502<br><br>Top chunks sent to LLM:<br>JEB0452-000453<br>JEB2554-224850<br>JEB1750-287502<br>JEB1161-191363<br>JEB0757-193353 | Jeb Bush expressed a supportive view of the H1B visa program during his tenure as governor of Florida. He recognized the importance of allowing skilled foreign workers, such as engineers, chemists, and physicists, to come to the U.S. to fill jobs that were difficult to fill domestically. He believed that limiting the number of H1B visas was a mistake and called on Congress to raise the cap to help maintain America's competitiveness and innovation. Bush emphasized the need for a flexible economy and the importance of having the right skill sets to fill jobs, acknowledging that while the goal is for Americans to fill every job, the reality required welcoming skilled foreign workers through programs like H1B visas [doc3].<br><br>However, there was also a critical perspective on the H1B program's expansion, suggesting that it contributed to a decline in tech jobs for U.S. citizens and created an uneven playing field, with concerns about the economic impact on American workers [doc1]. This indicates that while Bush saw the value in the program, there were nuanced views on its effects.<br><br>Citations:<br>[doc1] : JEB0088-123105<br>[doc3] : JEB0267-260436<br><br>Top chunks sent to LLM:<br>JEB0088-123105<br>JEB0882-015480<br>JEB0267-260436<br>JEB0221-146491<br>JEB1759-042509 |

| Prompt | Keyword | Vector | Keyword Semantic | Hybrid | Hybrid Semantic |
|---|---|---|---|---|---|
| What were Jeb Bush's views on the H1B visa program during his tenure as governor of Florida? Please ensure that all responses reflect Jeb Bush's own views specifically, not those of other individuals or political figures. | The retrieved documents do not contain any direct statements or views from Jeb Bush specifically about the H1B visa program during his tenure as governor of Florida. The documents include an email exchange where Jeb Bush responds to a request for assistance regarding an individual's H1B visa case, but he does not express his personal views on the H1B visa program itself [doc1][doc2]. Therefore, there is no available information in the provided documents that reflects Jeb Bush's own views on the H1B visa program.<br><br>Citations:<br>[doc1] : JEB1749-287501<br>[doc2] : JEB1750-287502<br><br>Top chunks sent to LLM:<br>JEB1749-287501<br>JEB1750-287502<br>JEB1311-157152<br>JEB3009-263178<br>JEB3052-263221 | The retrieved documents do not contain any information regarding Jeb Bush's views on the H1B visa program during his tenure as governor of Florida. Therefore, I am unable to provide an answer based on the provided data.<br><br>The requested information is not available in the retrieved data. Please try another query or topic.<br><br>Citations:<br><br>Top chunks sent to LLM:<br>JEB1161-191363<br>JEB0757-193353<br>JEB0333-098113<br>JEB0156-058019<br>JEB1503-086372 | Jeb Bush expressed support for increasing the cap on H1B visas during his tenure. He believed that allowing more highly skilled engineers, chemists, and physicists from other countries to work in the U.S. was important for filling jobs that were difficult to fill domestically. Bush emphasized that while the goal was to have Americans fill every job, the reality required being wise about letting skilled foreign workers come to the U.S. to help maintain America's competitiveness and leadership in innovation. He called upon Congress to be realistic and reasonable in raising the H1B visa cap to support this objective [doc1].<br><br>Citations:<br>[doc1] : JEB0267-260436<br><br>Top chunks sent to LLM:<br>JEB0267-260436<br>JEB0221-146491<br>JEB1125-280030<br>JEB3344-225640<br>JEB1870-131535 | The retrieved documents do not contain any direct statements or explicit views from Jeb Bush himself regarding the H1B visa program during his tenure as governor of Florida. There is an email exchange where Jeb Bush responds to a constituent's inquiry about an H1B visa case, indicating he would ask lawyers to look into it, but this does not provide a clear expression of his overall views on the H1B visa program [doc3].<br><br>Therefore, based on the available documents, Jeb Bush's specific views on the H1B visa program during his governorship are not documented.<br><br>If you need information on other topics or a different aspect of Jeb Bush's policies, please let me know.<br><br>Citations:<br>[doc3] : JEB1750-287502<br><br>Top chunks sent to LLM:<br>JEB0452-000453<br>JEB2554-224850<br>JEB1750-287502<br>JEB1161-191363<br>JEB0757-193353 | Jeb Bush expressed support for increasing the cap on H1B visas, emphasizing the importance of allowing more skilled foreign workers, such as engineers, chemists, and physicists, to fill jobs in the U.S. that are difficult to fill domestically. He believed that encouraging highly skilled individuals to work in America was essential for maintaining competitiveness, innovation, and a strong job base. Bush acknowledged the goal of having Americans fill every job but recognized the reality of the labor market and called on Congress to be realistic and raise the H1B visa cap to support economic flexibility and innovation [doc3].<br><br>However, there is also a contrasting critical view of the H1B program mentioned in a letter addressed to him, which he did not author, that links the expansion of the H1B program to negative impacts on the U.S. tech industry and employment for citizens. This critical perspective is not attributed to Jeb Bush himself but rather to the letter writer [doc1].<br><br>Therefore, Jeb Bush's own view was optimistic about the H1B visa program as a necessary tool to attract skilled workers to support the U.S. economy and innovation [doc3].<br><br>Citations:<br>[doc1] : JEB0088-123105<br>[doc3] : JEB0267-260436<br><br>Top chunks sent to LLM:<br>JEB0088-123105<br>JEB0882-015480<br>JEB0267-260436 |
| What were Jeb Bush's views on the H1B visa program during his tenure as governor of Florida? Please ensure that all responses reflect Jeb Bush's own views specifically, not those of other individuals or political figures, including President Bush. | The retrieved documents do not contain any information about Jeb Bush's views on the H1B visa program during his tenure as governor of Florida. Therefore, I cannot provide an answer based on the available data.<br><br>The requested information is not available in the retrieved data. Please try another query or topic.<br><br>Citations:<br><br>Top chunks sent to LLM:<br>JEB1035-199410<br>JEB0452-000453<br>JEB1039-081180<br>JEB3344-225640<br>JEB1166-193762 | The retrieved documents do not contain any information about Jeb Bush's views on the H1B visa program during his tenure as governor of Florida. Therefore, I cannot provide an answer based on the available data.<br><br>The requested information is not available in the retrieved data. Please try another query or topic.<br><br>Citations:<br><br>Top chunks sent to LLM:<br>JEB0333-098113<br>JEB0156-058019<br>JEB1503-086372<br>JEB0563-289391<br>JEB2554-224850 | The retrieved documents do not contain any information specifically about Jeb Bush's views on the H1B visa program during his tenure as governor of Florida. Therefore, I cannot provide an answer based on the available data.<br><br>The requested information is not available in the retrieved data. Please try another query or topic.<br><br>Citations:<br><br>Top chunks sent to LLM:<br>JEB0030-035501<br>JEB2721-276946<br>JEB0221-146491<br>JEB1642-037113<br>JEB1627-037098 | The retrieved documents do not contain any information about Jeb Bush's views on the H1B visa program during his tenure as governor of Florida. Therefore, I cannot provide an answer based on the available data.<br><br>The requested information is not available in the retrieved data. Please try another query or topic.<br><br>Citations:<br><br>Top chunks sent to LLM:<br>JEB0452-000453<br>JEB0563-289391<br>JEB1759-042509<br>JEB1164-193760<br>JEB0333-098113 | The retrieved documents do not contain any information about Jeb Bush's views on the H1B visa program during his tenure as governor of Florida. Therefore, I cannot provide an answer based on the available data.<br><br>The requested information is not available in the retrieved data. Please try another query or topic.<br><br>Citations:<br><br>Top chunks sent to LLM:<br>JEB0563-289391<br>JEB1759-042509<br>JEB1627-037098<br>JEB0947-275172<br>JEB0891-001711 |